\documentstyle[aps,multicol,psfig]{revtex}
\input epsf
\begin{document} 
%----------------------------
\newcommand{\omr}{\Omega_{r}}
\newcommand{\omp}{\Omega_{\phi}}
\newcommand{\omz}{\Omega_{z}}
\newcommand{\otil}{\tilde}
\newcommand{\ort}{\otil{\Omega}_{r}}
\newcommand{\opt}{\otil{\Omega}_{\phi}}
\newcommand{\ozt}{\otil{\Omega}_{z}}
\newcommand{\nhat}{\hat{N}}
\newcommand{\ntil}{\tilde{N}}
\newcommand{\mut}{\tilde{\mu}}
\newcommand{\arccot}{{\rm cot}^{-1}}
%----------------------------
\draft 
\title{Mesoscopic phenomena in Bose-Einstein systems: \\ 
Persistent currents, population oscillations and quantal phases}
%\columnsep .375in 
%\twocolumn[ 
\author{Yuli Lyanda-Geller and Paul M.~Goldbart} 
\address{Department of Physics and Materials Research Laboratory, \\ 
University of Illinois at Urbana-Champaign, Urbana, Illinois 61801, U.S.A.}
%----------------------------
%\date{\today}
 \date{December 8, 1998}
%----------------------------
\maketitle
%----------------------------
\begin{abstract} 
%----------------------------
Mesoscopic phenomena---including population oscillations and persistent 
currents driven by quantal phases---are explored theoretically in the 
context of multiply-connected Bose-Einstein systems composed of trapped 
alkali-metal gas atoms.  These atomic phenomena are bosonic analogues 
of electronic persistent currents in normal metals and Little-Parks 
oscillations in superconductors.
%----------------------------
\end{abstract}
%---------------------------- 
%----------------------------
%----------------------------
%-------------
% PACS numbers
%
% 03.75.Fi Phase coherent atomic ensemble (Bose condensation)
% 73.23.Ra Mesoscopic systems: Persistent currents
% 03.65.Bz Quantum mechanics: Foundations,... 
%	  (including Aharonov-Bohm, ..., Berry's phase)
% 74.25.Bt Superconductivity: Thermodynamic properties
%
% http://www.aip.org/pacs/
%----------
%----------
\begin{multicols}{2}
\narrowtext
%----------------------------
%----------------------------
\noindent
{\sl Introduction\/}: 
%----------------------------
The purpose of this Letter is to consider multiply-connected
many-particle systems obeying Bose-Einstein statistics and, in
particular, to address the sensitivity of such systems to quantal
phases.  If the bosonic constituents (e.g., alkali-metal gas atoms) are
electrically neutral, electromagnetism, in the form of the Aharonov-Bohm
(AB) phase, cannot provide a source of quantal phase.  Then, in seeking
sensitivity to quantal phases we are led to consider the spin degree of
freedom of the bosons, and the consequent possibility of quantal phases
of geometric origin~\cite{REF:Berry}. As we shall see, geometric quantal
phases can readily affect the energy levels, and hence populations, of
the single-particle quantum states, and lead to persistent equilibrium
currents in multiply-connected systems, thus providing a striking
example of quantal mesoscopic phenomena in the setting of bosonic
systems.  Such phenomena are bosonic analogues of phenomena well known
in the context of the mesoscopic physics of normal-state electronic
systems (such as persistent equilibrium currents and conductance
oscillations in conducting rings) which arise due to
AB~\cite{REF:AroShar} or geometric~\cite{REF:BPmeso} quantal phases.
They are also bosonic analogues of the flux-sensitivity of the
superconducting transition temperature of a thin superconducting ring,
known as Little-Parks oscillations~\cite{REF:LandP}.  (These quantum
interference phenomena are mesoscopic, in the sense that they vanish in
the limit of large system-size.)

It is worth mentioning that there is a sense in which bosonic 
settings are preferable to electronic settings, if one wishes to 
observe implications of quantal phases in many-particle physics: in the 
fermionic case, the Pauli exclusion principle forces the occupation of 
many single-particle states, and there are strong cancellations between 
the effects of quantal phases on these states.  By contrast, Bose-Einstein 
statistics promote the significance of the single-particle ground state.  
In this sense then, bosonic systems tend to amplify mesoscopic effects, at 
least in comparison with fermionic systems. 

One scheme for introducing a geometric quantal phase is to have 
the bosons move through regions of space in which there is a 
spatially varying magnetic field to which the spins of the 
bosons are Zeeman-coupled.  Then, as discussed in 
Ref.~\cite{REF:TLHo} in the context of magnetic traps, the 
inhomogeneous magnetic field (if sufficiently strong) leads to a 
geometric vector potential ${\bf A}$ (and a corresponding geometric 
flux $\Phi$) which influences the orbital motion of the bosons, 
and does so in much the same way as the electromagnetic (AB) 
vector potential (and flux) influences the motion of electrically 
charged particles.

%--------------------------
\begin{figure}[hbt]
%\vskip-3.25truecm
 \vskip-3.75truecm
\epsfxsize=\columnwidth
%\epsfysize=4.5cm
%\centerline{\epsfbox{FIGS/torus.eps}} 
 \centerline{\epsfbox{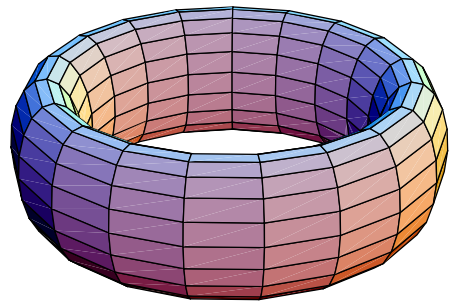}}
%\vskip-3.5truecm
 \vskip-3.75truecm
\caption{Toroidal sample of non-circular cross-section.}
\label{FIG:torus}
\end{figure}%
%--------------------------
References~\cite{REF:TLHo} considered conventional (i.e., not AB-like)
consequences of the geometric vector potential, i.e., effects associated
with nonzero values of the geometric field-strength 
${\bf\Omega}\equiv{\bf\nabla}\times {\bf A}$ 
(i.e., the vorticity).  However---and this is the main point of the
present Letter---there are striking quantal AB-like
consequences of the geometric vector potential itself (rather than the
field-strength), especially in multiply-connected configurations, and
even when ${\bf\Omega}={\bf 0}$ in the sample.  These consequences include 
an oscillatory dependence on the geometric flux $\Phi$ of the energies of 
all single-particle levels and, thus, the equilibrium populations of these
levels and, more generally, all equilibrium quantities.  Moreover, these 
single-particle energy-level oscillations lead directly to the existence
of equilibrium currents~\cite{REF:noneq} that flow around the trap (at
generic values of $\Phi$). 
Therefore, despite the electrical neutrality of the atoms in BEC
systems, the geometric phase allows one to realize counterparts of the
well-established collection of electromagnetic AB-phase sensitive
phenomena, as well as new, bosonic, phenomena such as large oscillations
in the populations of the various single-particle levels.

In order to realize geometric-phase--driven oscillations one needs to
vary the inhomogeneous magnetic field~\cite{REF:electric} that the
bosons inhabit.  In magnetic traps, the ability to make such variations
is limited, as variations alter the structure of the trap (i.e., the
shape of the system).  The recent achievement of confinement via purely
optical traps~\cite{REF:DMSK} liberates the magnetic field from its dual
role of confining the atoms {\it and\/} causing the geometric phase, and
thus enlarges the scope for exploring the effects of geometric phases.
We shall touch upon the issue of making multiply-connected traps at the
end of this Letter.

%----------------------------
\noindent
{\sl Model system\/}: 
%----------------------------
We consider a system of many identical charge-neutral noninterating bosonic 
atoms of mass $m$, confined to a multiply-connected trap.  For the sake of 
simplicity, we envisage the trap as having the following features: 
(i)~it is toroidal and axisymmetric; 
(ii)~it is sufficiently narrow in the radial direction that, under 
operating conditions, only the state with the lowest radial quantum number 
is occupied (i.e., the radial energy scale $\hbar\omr$ is large); 
(iii)~in the axial direction there is confinement by an oscillator 
potential; and 
(iv)~if the trap is optical then only mild conditions need be obeyed 
if an applied inhomogeneous magnetic field is not to change the 
trapping potential appreciably. 
(v)~If the trap is magnetic then the application of an additional 
homogeneous field will change $\Phi$ but not the confining potential.  
In the absence of $\Phi$, the spectrum of single-particle 
energy eigenvalues ${\cal E}_{\ell,n}^{0}$ is given by 
%%%
%\begin{equation}
${\cal E}_{\ell,n}^{0}=
 \hbar\omz\,n
+\hbar\omp\,\ell^{2}$, 
%\end{equation}
where $\hbar\omz$ is the axial oscillator energy scale, $\hbar\omp$ is 
the azimuthal energy scale, we have omitted all zero-point energy 
contributions, and we do not allow for radial excitation.  The quantum 
numbers $\ell$ and $n$ range, respectively, over all integers and all 
non-negative integers. 

We now suppose that the toroidal system of trapped atoms is subjected to a 
magnetic field having the property that the orientation of the field varies 
across the trap~\cite{REF:oppose}.  Under conditions of adiabaticity 
(i.e., $\omp$ much smaller than the Larmor frequency of the spins) for the 
dynamics of the spins of the atoms, the dominant effect of the inhomogeneous 
magnetic field on the energy spectrum is to introduce a spin-dependent flux 
$\Phi$, so that the spectrum (for atoms with spin-projection lying parallel 
to the magnetic field) becomes 
\begin{equation}
{\cal E}_{\ell,n}=
 \hbar\omz\,n
+\hbar\omp\,(\ell-\Phi)^{2}. 
\end{equation}
(Atoms with spin-projection lying antiparallel to the magnetic field 
direction are not trapped.)\thinspace\  The origin of this flux 
is the Berry phase associated with the spin dynamics (see, e.g., 
Refs.~\cite{REF:BPmeso}).  Our aim is to compute the number of particles 
in the single-particle ground state as a function of the temperature $T$, 
the geometric flux $\Phi$, and the mean total number of particles $N$, and 
to do so via the grand canonical ensemble~\cite{REF:ensemble}.  To this 
end, we first compute the total number of particles $N$, as a function of 
$T$, $\Phi$ and the chemical potential $\mu$: 
\begin{mathletters}
\begin{eqnarray}
N(T,\Phi,\mu)
&=&
% \sum_{\ell=-\infty}^{\infty}\sum_{n=0}^{\infty}N_{\ell,n}, 
  \sum\nolimits_{\ell=-\infty}^{\infty}
  \sum\nolimits_{n=0}^{\infty}N_{\ell,n}, 
\label{EQ:collect}
\\
N_{\ell,n}
&\equiv&
\big\{{\rm e}^{({\cal E}_{\ell,n}-\mu)/k_{\rm B}T}-1\big\}^{-1}.
%{1\over{{\rm e}^{({\cal E}_{\ell,n}-\mu)/k_{\rm B}T}-1}}.
\label{EQ:SPpops}
\end{eqnarray}%
\end{mathletters}%
Next, we decompose this sum: 
$N=\ntil+\nhat$, where 
\begin{equation}
\ntil\equiv\sum\nolimits_{\ell=-\infty}^{\infty}N_{\ell,0},
\quad
%\quad{\rm and}\quad
\nhat\equiv\sum\nolimits_{\ell=-\infty}^{\infty}
      \sum\nolimits_{n=1}^{\infty}N_{\ell,n}.
\end{equation}
To compute $\ntil$, we use a variant of the contour integration 
technique described, e.g., in Refs.~\cite{REF:Ol_Fett}, which yields
\begin{eqnarray}
\ntil
&=&
{\pi\over{2\varphi}}
\left\{\cot\pi\left(\Phi-\varphi\right)
     -\cot\pi\left(\Phi+\varphi\right)\right\}
\nonumber
\\
&&\qquad\qquad\qquad
+{\cal O}\Big(\exp\left(-(2\pi)^{3/2}/\opt^{1/2}\right)\Big), 
\label{EQ:ntilasy}
\end{eqnarray}
where we have introduced the reduced frequency 
$\opt\equiv\hbar\omp/k_{\rm B}T$, and the reduced 
chemical potential $\mut\equiv\mu/k_{\rm B}T$ and, for 
convenience, 
% $\varphi\equiv\sqrt{\mut/\opt}$.
$\varphi$ denotes $\big({\mut/\opt}\big){^{1/2}}$.
[The form for $\ntil$ arrived at by this technique is much more rapidly 
convergent than the original form, Eq.~(\ref{EQ:collect})]. 
Not surprisingly, however, even after omitting exponentially small 
terms, the transcendental equation for $\mut(T,\Phi,\ntil)$, 
Eq.~(\ref{EQ:ntilasy}), cannot, in general, be solved explicitly. 
Without loss of generality, let us assume that $0\le\Phi\le 1/2$.  
(Results for other values of $\Phi$ can be obtained via the symmetries 
of reflection, $\Phi\to-\Phi$, and translation, $\Phi\to\Phi+1$.) 
Then further simplification is achievable in three cases: 
(i)~$\Phi\ll 1/2$, 
(ii)~$\Phi\gg\big({\ntil\opt}\big){^{-1/2}}$, and 
(iii)~$\vert\cot 2\pi\Phi\vert\gg\ntil\opt\Phi/\pi$.
In case~(i) we expand the cotangents on the r.h.s~of 
Eq.~(\ref{EQ:ntilasy}) in Laurent series, 
retaining two terms in each series, and solve the resulting 
equation for $\mut$, thus arriving at 
\begin{equation}
\mut\approx
\opt\Phi^{2}-
% {1\over{\ntil-\big(\pi/3\opt\big)}}.
\{\ntil-\big(\pi/3\opt\big)\}^{-1}.
\label{EQ:mutilasy}
\end{equation}
In case~(ii) we instead expand the prefactor and the arguments of the 
cotangents in Eq.~(\ref{EQ:ntilasy}) to linear order in the deviation 
of $\mut$ from the (dimensionless) single-particle ground-state energy 
$\opt\Phi^{2}$.  Thus, we arrive at 
\begin{eqnarray}
\mut
&\approx&
\opt\Phi^{2}-
2\pi^{-1}\Phi\,
\arccot
\Big(
 \cot 2\pi\Phi
+\pi^{-1}\Phi\opt\ntil
%\right.
\nonumber
\\
&&\qquad\qquad
%\left.
+\sqrt{\cot^{2}\pi\Phi
+(\Phi\opt\ntil/\pi)^{2}+1}
 \,\,\Big).
\label{EQ:mutiltwo}
\end{eqnarray}
In case~(iii), which corresponds to $\Phi$ close to 1/2, we have
\begin{equation}
\mut\approx
\opt\Phi^{2}-
2\pi^{-1}\Phi\,\arccot
\big(\Phi\opt\ntil/\pi\big). 
\label{EQ:mutilthree}
\end{equation}
In making our expansions of Eq.~(\ref{EQ:ntilasy}) for $\ntil$ we
restrict ourselves to the regime of quasi-BEC, i.e., we consider values
of $\mu$ only slightly smaller than the single-particle ground-state
energy ${\cal E}_{0,0}$ (i.e., $\hbar\omp\Phi^{2}$).

%----------------------------
\noindent
{\sl Physical consequences of the geometric flux\/}: 
%----------------------------
To use these results for $\mut(T,\Phi,\ntil)$ to determine desired physical 
quantities, such as the populations $N_{\ell,n}$ of the single-particle 
states as functions of the variables $(T,\Phi,N)$, we first note that in 
the regime of BEC we need only retain the difference between $\mu$ and 
${\cal E}_{0,0}$ in $\ntil$, but may omit 
%--------------------------
\begin{figure}[hbt]
\vskip-0.25cm
\epsfxsize=\columnwidth
\epsfysize=5.5cm
%\epsfxsize=5.0truecm 
%\centerline{\epsfbox{FIGS/population.eps}} 
 \centerline{\epsfbox{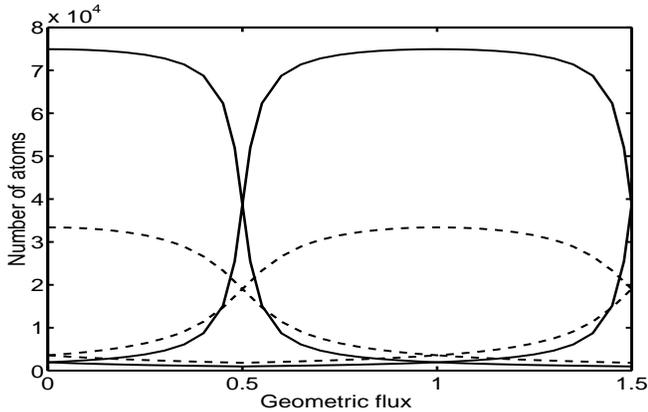}}
\vskip+0.4truecm
\caption{Oscillatory dependence on the geometric flux of the 
populations of the lowest three single-particle energy 
levels at a higher (dashed line) and a lower (full line) 
temperature.}
\label{FIG:population}
\end{figure}%
%----------------------------
\noindent
it from $\nhat$. 
Next, we observe that 
$N\equiv\ntil+\nhat$, so that by knowing $\mut(T,\Phi,\ntil)$ we 
know $\mut(T,\Phi,N)$.  This we use to eliminate $\mu$ from the 
Bose functions that determine the populations $N_{\ell,n}$.
To illustrate the oscillatory behavior of single-particle state 
populations with $\Phi$ we show, in Fig.~\ref{FIG:population}, 
the populations $N_{\ell,m}$ of the three lowest-lying states 
(i.e., $N_{0,0}$, $N_{1,0}$ and $N_{2,0}$, as $\vert\Phi\vert\le 1/2$)
as functions of $\Phi$~\cite{REF:numerics}.  
%----------------------------
We have chosen for this illustration a system of 
$N=10^{5}$ atoms of $^{87}$Rb, 
sample radius $R=1\,\mu{\rm m}$, 
axial trap frequency $\omz=7,500\,{\rm Hz}$, 
temperature 
$T=5\,\mu{\rm K}$  (for which $800\opt=40\ozt=1$) or 
$T=10\,\mu{\rm K}$ (for which $400\opt=20\ozt=1$),  
where $\ozt{\equiv}\hbar\omz/k_{\rm B}T$.  
%----------------------------
Note that at $\Phi{=}0$ we have $N_{\ell,n}{=}N_{-\ell n}$, and at
$\Phi{=}1/2$ we have $N_{\ell,n}{=}N_{\ell+1,n}$.  The latter case
illustrates the more general point that at half-integral values of
$\Phi$ the lowest single-particle energy level is degenerate for the
case of perfectly azimuthally symmetric traps.  If, however, the
azimuthal symmetry is absent then the level crossing is avoided, and the
single-particle ground state is separated from the excited states by an
energy gap at all values of $\Phi$.  

The population oscillations are
mesoscopic, in the sense that they vanish in the thermodynamic
limit~\cite{REF:tdl}.  Indeed, the number of atoms in traps, although
typically large, is not on the order of Avogadro's number and,
therefore, the systems are even further from the thermodynamic limit
than conventional macroscopic and even mesoscopic samples. Nevertheless,
although BEC is not, strictly speaking, marked by a sharp thermodynamic
phase transition, atomic condensates acquire features of the
thermodynamic limit already at $N=10^{4}$ (see, e.g.,
Refs.~\cite{REF:KettDru,REF:pit}). Despite this, our calculations reveal
oscillatory phenomena at $N=10^{5}$ whenever the fraction of atoms in
the ground state is appreciable.  Thus,  one has the capability of
observing, simultaneously, both macroscopic and mesoscopic phenomena.

The $\Phi$-dependence of the single-particle energy levels also
leads to the phenomenon of equilibrium persistent currents, i.e., 
dissipationless particle-currents that flow around the trap.  (Such
equilibrium currents should, of course, be distinguished from the
nonequilibrium metastable currents that can arise in multiply-connected
samples; see, e.g., Refs.~\cite{REF:noneq}.)\thinspace\ The
simple formula, 
\begin{equation}
I=\sum\nolimits_{\ell=-\infty}^{\infty}
  \sum\nolimits_{n=0}^{\infty}
N_{\ell,n}\,\partial{\cal E}_{\ell,n}/\partial\Phi, 
\label{EQ:persist}
\end{equation}
%--------------------------
\begin{figure}[hbt]
\vskip-0.5truecm
\epsfxsize=\columnwidth
\epsfysize=4.0cm
%\epsfxsize=5.0truecm 
%\centerline{\epsfbox{FIGS/pers_cur.eps}} 
 \centerline{\epsfbox{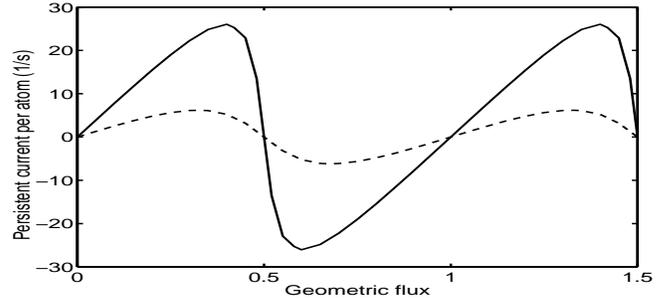}}
\vskip+0.4truecm
\caption{Oscillatory dependence of the equilibrium persistent 
particle-current (per atom) on the geometric flux at a higher 
(dashed line) and a lower (full line) temperature.}
\label{FIG:persistent}
\end{figure}%
%--------------------------
\noindent
for the persistent particle-current $I$ follows from the general
relation $\delta F/\delta {\bf A}=-{\bf j}$, where $F$ is the free
energy, ${\bf A}$ is a gauge potential (such as the geometric vector
potential) and ${\bf j}$ is the conjugate current-density.  In
Fig.~\ref{FIG:persistent} we show the dependence of $I/N$ on $\Phi$ 
at the conditions and temperatures specified in the previous paragraph.  
As $T$ is increased from zero the saw-tooth form of the $\Phi$-dependence 
is smoothed to a more sinusoidal form, but the zeros at integral and 
half-integral values of the flux are preserved.  As for the 
characteristic scale of the amplitude of $I$, it is on the order of 
$N\omp$, at least when the ground state contains most of the particles.

%--------------------------
\begin{figure}[hbt]
\vskip-0.25cm
\epsfxsize=\columnwidth
\epsfysize=4.0cm
%\epsfxsize=5.0truecm 
%\centerline{\epsfbox{FIGS/spec_heat.eps}} 
 \centerline{\epsfbox{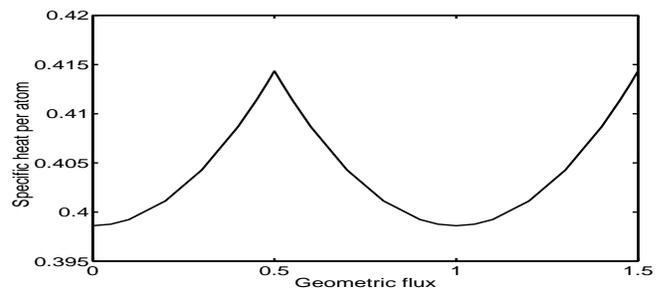}}
\vskip+0.4truecm
\caption{Oscillatory dependence of the specific 
heat per atom on the geometric flux.}
\label{FIG:specific}
\end{figure}%
%--------------------------
As it is Fermi rather than Bose systems that have traditionally 
provided settings for mesoscopic physics, we 
pause to compare the characteristic magnitudes of persistent 
currents in Fermi and Bose systems.  In the special case of 
single-channel systems the characteristic magnitudes are similar: 
the many bosons in the (low-velocity) ground state contributing 
roughly as much as the single (high-velocity) fermion at the Fermi 
level (contributions from fermions below the Fermi level 
essentially cancelling one another).  In the more general case, 
however, in which there are many channels, the greater the extent 
of transverse excitation, the smaller the contribution to the 
persistent fermion current (owing to the reduced kinetic energy 
at the Fermi level).  By contrast, for bosons the particle 
occupations are, of course, not spread over the many current-reduced 
channels, and instead are concentrated on the optimal channel. 
Thus, for bosonic systems such mesoscopic effects are amplified, 
relative to the Fermi case.

As a third consequence of the geometric flux, we consider the 
oscillatory behavior of the (dimensionless) specific heat (per particle) 
$C\equiv\partial E(T,\Phi,N)/\partial Nk_{\rm B}T$.  (We recognize that 
$C$ may be difficult to measure.)\thinspace\  For $C$, a more pronounced 
$\Phi$-dependence arises in the case of traps that are weakly confining in 
the axial direction (i.e., $\omz\ll\omp$).  In Fig.~\ref{FIG:specific} we 
show $C(\Phi)$ for the case of $^{23}$Na (the reduced 
mass of which also enhances the sensitivity to $\Phi$ compared with 
$^{87}$Rb) at $R=0.5\,\mu{\rm m}$, $T=40\,{\rm nK}$ and $\omz=100\,{\rm Hz}$.  
Upon closer inspection, the apparent cusp at the level crossing, which is  
a remnant of the true singularity at $T=0$, is seen to be rounded.

%----------------------------
\noindent
{\sl Experimental issues and concluding remarks\/}: 
%----------------------------
Having described several consequences of the geometric flux, we now briefly 
discuss some issues concerning the possibility of the experimental 
realization of these consequences.  
%----------------------------
We see three pivotal matters: 
  (i)~how to construct a toroidal sample; 
 (ii)~how to subject the sample to a suitably inhomogeneous field; and 
(iii)~how to detect population oscillations and persistent currents. 
%----------------------------
As for~(i), it should be feasible to construct toroidal samples based 
on magnetic traps by using a blue-detuned laser to repel atoms from 
the trap center.  We hope that in purely optical traps~\cite{REF:DMSK} 
a similar method, combining red- and blue-detuned lasers, 
could be employed to make a toroidal sample. 
%----------------------------
As for~(ii), magnetic trap technology itself is suitable for creating 
the necessary textured magnetic fields.  (As magnetic traps are 
currently used as a first stage in the loading of optical traps, 
creating textured fields should not impose a large additional 
experimental burden.)\thinspace\ 
%----------------------------
As for~(iii), the phonon-imaging technique~\cite{REF:sound} discussed
with regard to metastable currents in the last paragraph of the second  
of Refs.~\cite{REF:noneq} is less difficult in the present
setting of equilibrium persistent currents, owing to the far larger
magnitude of the latter.  Other experimental possibilities may include
(nondestructive) light-scattering and (destructive) time-of-flight
techniques.

We note that interactions between the bosons are not expected to alter 
qualitatively the oscillatory effects that we have discussed.  This is 
because, in the present context, $k_{\rm B}T\gg\mu$, for which one expects 
interactions to have only perturbative consequences~\cite{REF:pit}.  We 
conclude by remarking that if it is possible to realize superconductivity 
via the BEC of pre-formed bosons then oscillations in the level-populations, 
persistent currents, and perhaps specific heat, similar to those described 
in the present Letter, may be observable as consequences of an AB flux.

%----------------------------
\noindent
% {\sl Acknowledgments\/}: 
%----------------------------
Useful discussions with
A.~Bala{\-}eff, C.~Bender, A.~J.~Leg{\-}gett
and the UIUC/CNRS Bose-Einstein Condensation 
Workshop participants are gratefully acknowledged.
This work was supported by the Department of Energy, 
Division of Materials Sciences, Grant~DEFG02-96ER45439.
%-----------------------------
%----------------------------
  
%----------------------------
%----------------------------
\end{multicols}

\begin{references}
\bibitem{REF:oppose}
In optical traps one can use Helmholtz coils to create radial fields. 
By adding a homogeneous field one can create, e.g., crown-shaped fields.
\bibitem{REF:ensemble}
Although it seems more appropriate to use the canonical (i.e., fixed 
particle-number) ensemble, it is more straightforward to use  
the grand-canonical ensemble, as we have done.  We hope that the error 
incurred is negligible, at least at the particle numbers on 
which we are focusing.
\bibitem{REF:numerics}
For this illustration we have computed the populations numerically,
using Eqs.~(\ref{EQ:collect}) and (\ref{EQ:SPpops}).
\bibitem{REF:tdl}
By the thermodynamic limit we mean 
$N\to\infty$, $\omp$ and $\omz\to 0$, but with 
% $N\omr\sqrt{\omp}$ 
  $N\omr\omp^{1/2}$ 
tending to a constant.  In this limit, we arrive at a system 
that, although two-dimensional geometrically, does indeed 
exhibit thermodynamically sharp BEC, owing to the precise 
form of the density of states associated with the axial 
oscillator.
\end{references}
\end{document}